\documentclass{ifacconf}
\usepackage{natbib}   

\usepackage{graphicx}     
\usepackage{amsmath}
\usepackage{tikz}
\usepackage{interval}
\usepackage{amssymb}
\usepackage{placeins}

\newcommand{\Dks}{D_\text{KS}}
\newcommand{\barDks}{\bar{D}_\text{KS}}
\newcommand{\W}{\mathcal{W}}
\newcommand{\U}{\mathcal{U}}

\newcommand\figref{Fig.~\ref}
\newcommand\secref{Section~\ref}

\newcommand\tabref{Table~\ref}
\usepackage{booktabs}
\usepackage{amsmath}

\usepackage{bm}
\usepackage{times}

\makeatletter
\let\old@ssect\@ssect 
\makeatother

\usepackage{natbib}
\usepackage{hyperref}

\makeatletter
\def\@ssect#1#2#3#4#5#6{%
  \NR@gettitle{#6}
  \old@ssect{#1}{#2}{#3}{#4}{#5}{#6}
}
\makeatother

\begin{document}
\begin{frontmatter}

\title{Energy-Based Survival Models for Predictive Maintenance} 

\thanks[footnoteinfo]{This work has been submitted to IFAC for possible publication.}

\author[First]{Olov Holmer} 
\author[First]{Erik Frisk}
\author[First]{Mattias Krysander} 

\address[First]{Department of Electrical Engineering, Linköping University,
SE-58183 Linköping, Sweden, 
   (e-mail: name.lastname@liu.se).}

\begin{abstract}                

   Predictive maintenance is an effective tool for reducing maintenance costs. Its effectiveness relies heavily on the ability to predict the future state of health of the system, and for this survival models have shown to be very useful. Due to the complex behavior of system degradation, data-driven methods are often preferred, and neural network-based methods have been shown to perform particularly very well. Many neural network-based methods have been proposed and successfully applied to many problems. However, most models rely on assumptions that often are quite restrictive and there is an interest to find more expressive models. Energy-based models are promising candidates for this due to their successful use in other applications, which include natural language processing and computer vision. The focus of this work is therefore to investigate how energy-based models can be used for survival modeling and predictive maintenance.  A key step in using energy-based models for survival modeling is the introduction of right-censored data, which, based on a maximum likelihood approach, is shown to be a straightforward process.  Another important part of the model is the evaluation of the integral used to normalize the modeled probability density function, and it is shown how this can be done efficiently. The energy-based survival model is evaluated using both simulated data and experimental data in the form of starter battery failures from a fleet of vehicles, and its performance is found to be highly competitive compared to existing models.

\end{abstract}


\end{frontmatter}

\section{Introduction}
In many industrial applications, low maintenance costs and high uptime is essential, and for this predictive maintenance has shown to be an important technique. Predictive maintenance aims to determine when maintenance should be done based on observations of the system. For this, predictions of how the system evolves over time are needed, and by estimating the lifetime distribution of the system survival modeling has shown to be an effective tool. Due to the complex behavior of system degradation, data-driven methods are attractive and, while other methods like random survival forests \citep{ishwaranRandomSurvivalForests2008} exist, methods based on neural networks have shown to be superior for this task. 

Numerous neural network-based survival models have been proposed, see for example \cite{liAttentionbasedDeepSurvival2022} for a review. Many of the available methods rely on the assumption of proportional hazards \citep{coxAnalysisMultivariateBinary1972} where the scaling factor is taken as the output from a neural network, see for example \cite{chingCoxnnetArtificialNeural2018} and \cite{katzmanDeepSurvPersonalizedTreatment2018}. Similarly, methods where a specific distribution is assumed, and a neural network is used to predict the parameters of the distribution have also been proposed, like in \cite{dhadaWeibullRecurrentNeural2022}. While these have shown to be very useful approaches in many cases, the assumptions are very restrictive and often not sensible.

Another common approach is to model a discrete-time distribution where the time is discretized on a grid and the probability of failure at each time in the grid is given by the network, see for example \cite{brownUseArtificialNeural1997}, \cite{biganzoliFeedForwardNeural1998}, and \cite{gensheimerScalableDiscretetimeSurvival2019}. 
 These methods can also be extended to continuous-time by interpolating the discrete-time distribution, as shown in \citep{kvammeContinuousDiscretetimeSurvival2021} and \citep{voronovPredictiveMaintenanceLeadacid2020}, the training is however still done in discrete-time.
A benefit of these methods is that they do not rely on any assumptions regarding the distribution, other than that they are formulated in discrete time, and by using a small enough grid size they can theoretically become as expressive as desired. In practice, however, the grid size is limited by the amount of data, as shown in \cite{kvammeContinuousDiscretetimeSurvival2021}. Their results also indicate that discretization has a larger impact on performance than the choice of method. 


The methods above all rely on assumptions that often are quite restrictive and exploring methods that can utilize more of the expressiveness of neural networks is therefore of interest. For this, energy-based models \citep{lecunTutorialEnergyBasedLearning} are candidates. They can be used to specify a probability density directly via a neural network with a scalar output, making them highly expressive. They have been used successfully in many applications, including natural language processing \citep{bengioNeuralProbabilisticLanguage2000} and computer vision \citep{duImplicitGenerationModeling2019,gaoLearningGenerativeConvnets2018}, and it is interesting to see how these results translate to survival modeling.

With the overall goal of exploring more expressive neural network-based survival models, the focus of this work is to investigate how energy-based models can be used for survival modeling and predictive maintenance.


\section{Survival Modeling for Predictive Maintenance}
In this work, we consider the task of replacing a specific component in a system before it fails. If the component is not changed in time this could lead to unwanted downtime which can be costly. At the same time, changing the component too often is also costly. To handle this trade-off survival models can be used, and the remainder of this section provides the foundations of survival modeling and how it can be used for predictive maintenance.

\subsection{Survival Modeling}
Survival analysis is the statistical analysis of the duration of time until a specific event occurs, in this case the failure time $T$ of the component of interest. Survival models are often described using the survival function, which describes the probability that a component survives $t$ time units, and is defined as
\begin{equation}
   S(t,x) = p(T>t\mid x) = \int_t^\infty f(\tau, x)\, d\tau
\end{equation}
where $f(t,x)$ is the failure probability density function and $x$ is the covariate vector. The goal in survival modeling is to find a model $S_\theta$, with parameters $\theta$, that predicts the true survival function based on the recorded data.

\subsubsection{Censoring}
Usually not all recorded events are the actual failure times; instead, some are the time when the individual dropped out due to some other reason, for example, the end of the experiment or some unconsidered failure in the system. This means that the data contains right-censoring, and this is a central problem in survival modeling that must be considered when predicting the survival function. This means that the survival data from $N$ subjects have the form $\left\{(x_i,\tau_i,\delta_i) \right\}_{i=1}^N$ where, for each subject $i$, $\tau_i$ is the time of the event, and $\delta_i=1$ and  $\delta_i=0$ corresponds to the event being a failure and a censoring time, respectively.

\subsubsection{Likelihood}
The de facto standard for fitting survival models is maximum likelihood estimation. This is done by maximizing the likelihood, which for a recorded failure is
\begin{equation}
   L(\theta\mid \tau_i, x_i, \delta_i = 1) = p(T = \tau_i \mid \theta ,x_i) = f_\theta(\tau_i,x_i)
\end{equation}
and for a censored event
\begin{equation}
   L(\theta\mid \tau_i, x_i, \delta_i = 0) = P(T > \tau_i \mid \theta, x_i) = S_\theta(\tau_i, x_i).
\end{equation}
Based on this, the total likelihood becomes
\begin{equation}\label{eq:likelihood}
   L\left(\theta\mid \{(x_i,\tau_i,\delta_i)\}_{i=1}^N\right) = \prod_{i\mid \delta_i=1} f_\theta(\tau_i, x_i) \prod_{i\mid \delta_i=0} S_\theta(\tau_i, x_i)
\end{equation}
where the products are taken over the sets of $i$ where $\delta_i=1$ and $\delta_i=0$, respectively. In practice, however, the logarithm of the likelihood is typically used.
 
\subsection{Survival Model-Based Predictive Maintenance}

Based on the information of the system up to some time $t_0$ the problem is to decide if the component should be changed now or if it can be continued to be used for some time. Based on the predicted conditional survival function 
\begin{equation}\label{eq:csf}
   S^{t_0}_\theta(t,x) = p(T>t+t_0\mid T\geq t_0, x) = \frac{S_\theta(t,x)}{S_\theta(t_0,x)}
\end{equation}
a decision to change the component can be taken based on if 
\begin{equation}
   S^{t_0}_\theta(t_h,x) < \mathcal{J}
\end{equation}
where $\mathcal{J}$ is a threshold corresponding to the probability that the component survives the desired time horizon $t_h$. 

The conditional survival function can be calculated based on the quotient of the survival function as shown in \eqref{eq:csf}. However, we are not interested in the values of the survival function for $t<t_0$, and the meaning of 
\begin{equation}
   S(t,x) = p(T>t\mid x)
\end{equation}
for $t<t_0$ is not completely clear since $x$ contains information of the system from time $t=t_0$. For this reason, the conditional survival function is estimated directly by considering simplicity considering the duration of time from $t_0$ to the failure. 



\section{Energy-Based Models}

Energy-based models  are used to describe a probability density function defined, in terms of survival analysis, as
\begin{equation}
   p(t \mid x,\theta) = \frac{e^{-E_\theta(t,x)}}{Z_\theta(x)}
\end{equation}
where $E_\theta(t,x)$ is the output from a neural network, with parameters $\theta$, and is interpreted as the energy of the density, and 
\begin{equation}
   Z_\theta(x) = \int_0^\infty e^{-E_\theta(\tau,x)} \, d\tau
\end{equation}
is a normalization constant. The integral above is in general intractable which complicates its evaluation and is the main reason why energy-based models are often considered difficult to work with. This problem is left for now and is instead treated later in \secref{sec:integration}.

The failure density function can be taken directly as
\begin{equation}
   f_\theta(t,x) = \frac{e^{-E_\theta(t,x)}}{Z_\theta(x)},
\end{equation}
while the survival function requires an additional integration
\begin{equation}
   Z_\theta(t,x) = \int_t^\infty e^{-E_\theta(\tau,x)} \, d\tau
\end{equation}
before it can be calculated as
\begin{multline}
   S_\theta(t,x) = \int_t^\infty f_\theta(\tau, x) \,d\tau = \frac{1}{Z_\theta(x)}\int_t^\infty e^{-E_\theta(\tau,x)} \,d\tau \\=  \frac{Z_\theta(t,x)}{Z_\theta(x)}
\end{multline}

\subsection{Maximum Likelihood Training}
Many approaches to training energy-based models have been proposed, see for example \cite{songHowTrainYour2021}, and \cite{gustafssonHowTrainYour2020}. A common approach is to use maximum likelihood training where the normalization constant is evaluated using Monte Carlo methods. A benefit of this method is that it is straightforward to include censored data, as can be seen below. 

Maximum likelihood training is typically performed by minimizing the negative log-likelihood, which, based on \eqref{eq:likelihood} is
\begin{multline}
   J\left(\theta\mid \{(\tau_i,E_i)\}_{i=1}^N\right) = - \log L\left(\theta\mid \{(\tau_i,\delta_i)\}_{i=1}^N\right)   \\
   = -\sum_{\delta_i=1} \log f_\theta(\tau_i,x_i) - \sum_{\delta_j=0} \log S_\theta(\tau_j,x_j) \\
   = 
   \sum_{\delta_i=1} E_\theta(\tau_i,x_i) - \sum_{\delta_j=0} \log Z_\theta(\tau_j,x_j)  + \sum_k \log Z_\theta(x_k).
\end{multline}
The model can now be trained using standard machine learning techniques like gradient descent-based methods. However, the normalization constant must first be calculated by evaluating the integral $Z_\theta(x)$, and in the case of censoring the integral $Z_\theta(t,x)$ must also be evaluated.

\subsection{Evaluating the Integrals} \label{sec:integration}
\begin{figure}
   \begin{center}
      \includegraphics[width=\columnwidth]{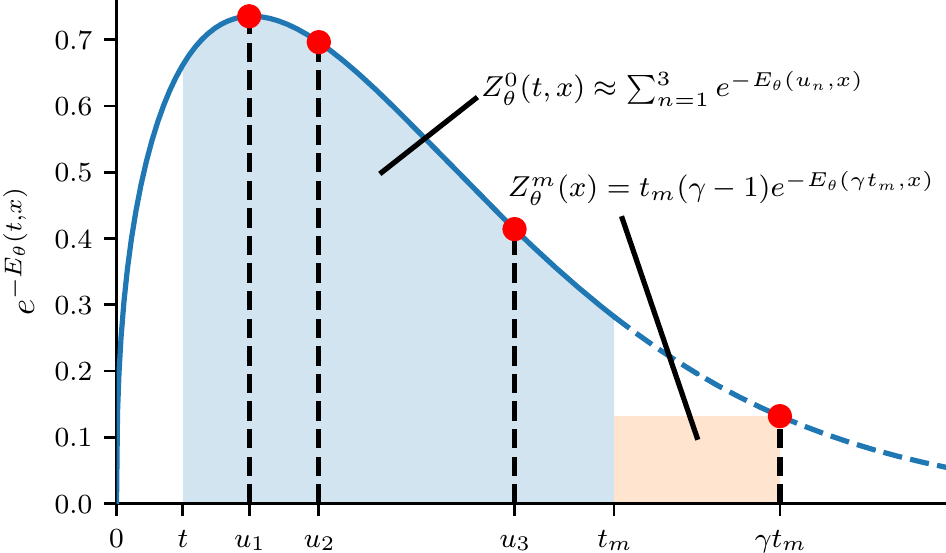}
   \end{center}
   \caption{Illustration of the integration when using three random samples. The full blue line indicates the part where the model is trained to represent the true distribution as close as possible, and in the dotted part, only the value at $\gamma t_m$ is of interest. This means that the ideal value of the integration is the sum of the blue and orange areas. The orange area is easily calculated, but the blue area is approximated using Monte Carlo integration. }\label{fig:itegration_illustration}

\end{figure}

The integral is typically evaluated using sampling-based Monte Carlo methods. One reason for this is that the integral contains infinite limits, which is also the case here. 
However, since survival data is usually recorded during a finite period, there exists a time $t_m$ after which no events are recorded. This means that the data contain limited information about the distribution after this time, in the sense that it only contains information about the probability of the failure happening after $t_m$ and not the shape of the distribution. For this reason, the integral is split into two parts
\begin{equation}
   Z_\theta(t,x) = Z_\theta^0(t,x) + Z_\theta^m(x)
\end{equation}
where 
\begin{equation}
   Z_\theta^0(t,x) = \int_t^{t_m} e^{-E_\theta(\tau,x)} \, d\tau
\end{equation}
and
\begin{equation}
   Z_\theta^m(x) = \int_{t_m}^\infty   e^{-E_\theta(\tau,x)} \, d\tau
\end{equation}
The latter corresponds to the tail of the distribution and since we are not interested in the shape of the tail its integral is approximated using a single point $\gamma t_m$ as
\begin{equation}
   Z_\theta^m(x) = \left( \gamma t_m -t_m \right) e^{-E_\theta(\gamma t_m,x)}
\end{equation}
where $\gamma>1$ is a constant. This can be interpreted as an implicit assumption that the survival function goes linearly to zero in the interval $\interval{t_m}{\gamma t_m}$. In \figref{fig:itegration_illustration} an illustration of the integration scheme is shown.

The remaining part $Z_\theta^0(t,x)$ is a proper integral allowing standard numerical methods to be used. However, if a fixed grid is used, only the values at the grid points are included in the integration and thus the normalization will not have the intended effect since the energy between two grid points can be made arbitrarily large without affecting the integral.  
For this reason, a sampling-based method is still used. While many sampling-based methods have been proposed, for simplicity and to establish a baseline, the Monte Carlo method using uniform sampling of the interval $\interval { 0 }{ t_m }$ is used. 

The problems of using a fixed grid are all limited to the training of the model, meaning that when the model is used for prediction fixed grids can be used. For this reason, the trapezoidal rule with a uniform grid is used during prediction.

\section{Data and Experiments}
To evaluate the performance of the energy-based survival model, two different datasets are used: one based on simulations and one based on starter battery failures from a fleet of vehicles.

\subsection{Simulated Data}
\begin{figure}
   \begin{center}
      \includegraphics[width=\columnwidth]{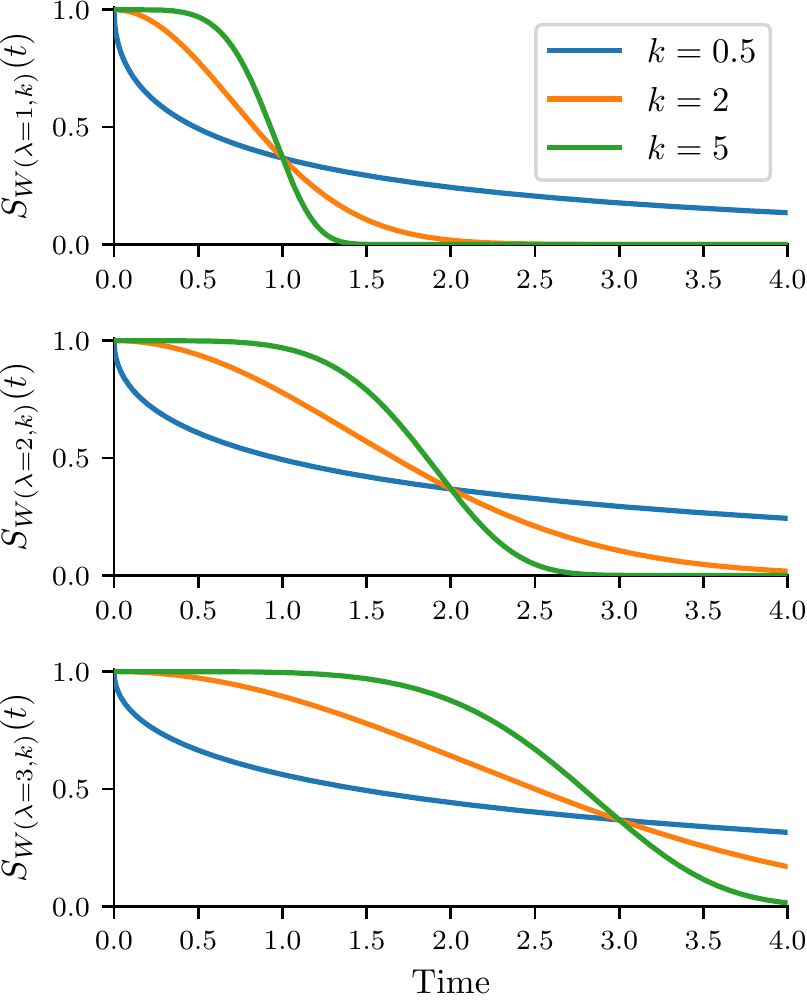}
   \end{center}
   \caption{The survival function of the simulated data for three values of $k$ and $\gamma$.}
   \label{fig:weibull}
\end{figure}

The simulated data is drawn from a two-parameter Weibull distribution with the density function
\begin{equation}
   f_{\W(\lambda,k)}(t) = \frac{k}{\lambda} \left( \frac{t}{\lambda} \right)^{k-1} e^{-\left( \frac{t}{\lambda} \right)^{k}}
\end{equation}
and survival function
\begin{equation}
   S_{\W(\lambda,k)}(t) = e^{-\left( \frac{t}{\lambda} \right)^{k}}
\end{equation}
where $k > 0$ is the shape parameter and $\lambda > 0$ is the timescale parameter of the distribution. For each individual $i$ the parameters are drawn from uniform distributions, according to $\lambda_i \sim \U(1,3)$ and $k_i \sim \U(0.5,5)$, and the covariate vector is taken as $x_i=[\lambda_i, k_i]$. In \figref{fig:weibull} the survival function for three values of $k$ and $\gamma$ is shown.

Censoring is introduced by also drawing a censoring time $C_i\sim \U(0,3)$, and taking the recorded time as $\tau_i = \min(T_i,C_i)$. This corresponds to a censoring rate of around 56~\%.

\subsubsection{Kolmogorov–Smirnov Distance}
One of the benefits of using simulated data is that the predicted survival function $S_\theta(t;\lambda,k)$ can be compared with the true $S_{\text{W}(\lambda,k)}(t)$. For this the Kolmogorov–Smirnov distance
\begin{equation}
   \Dks^{t_m} \left(S_\theta(\cdot;\lambda,k), S_{\W(\lambda,k)}\right) = \sup_{ t \leq t_m} \left| S_\theta(t;\lambda,k) - S_{\W(\lambda,k)}(t)  \right| 
\end{equation}
is used. The supremum is evaluated numerically by using an equidistant grid of 100 points between $0$ and  $t_m$. 

To evaluate the performance of the model on the whole population the mean Kolmogorov–Smirnov distance
\begin{multline}
   \barDks^{t_m}(S_\theta) = \mathbb{E}  \Dks \left(S_\theta(\cdot;\lambda,k), S_{\W(\lambda,k)}\right) \\\approx \frac{1}{N_\lambda N_k} \sum_{\lambda_n} \sum_{k_m}  \Dks \left(S_\theta(\cdot;\lambda_n,k_m), S_{\W(\lambda_n,k_m)}\right)
\end{multline}
where the sums are taken over $N_\lambda=N_k=20$ values of $\lambda$ and $k$ uniformly distributed in their respective range is used.

\subsection{Vehicle Fleet Data}
The vehicle fleet data consists of starter battery failure times from around 25,000 vehicles, with a censoring rate of 74~\%. The features, or covariates, in this data set consists of categorical variables specifying the type of vehicle and operational data from the first year of usage (i.e. $t_0 = 1$~year).  The operational data is based on signals from the vehicles' control systems, selected by experts based on the belief that they can be useful for predicting battery failures. In total there are 100 features in the data, all of which are normalized to be in the range $\interval{0}{1}$. The data is also split into three parts: a training set, a test set, and a validation set. The training set and validation set are used during the training, the training set to estimate the gradient and the validation set to monitor the progress. The test set is only used for independent evaluation of the trained model. 10~\% of the data is used for the validation set, 20~\% for the test set, and the remaining 70~\% for the training set.


\section{Results}
In this section, the results from applying the energy-based survival model on the simulated and experimental data are presented. 

\subsection{The Importance of the Number of Samples in the Integration}
\begin{figure}
   \begin{center}
      \includegraphics[width=\columnwidth]{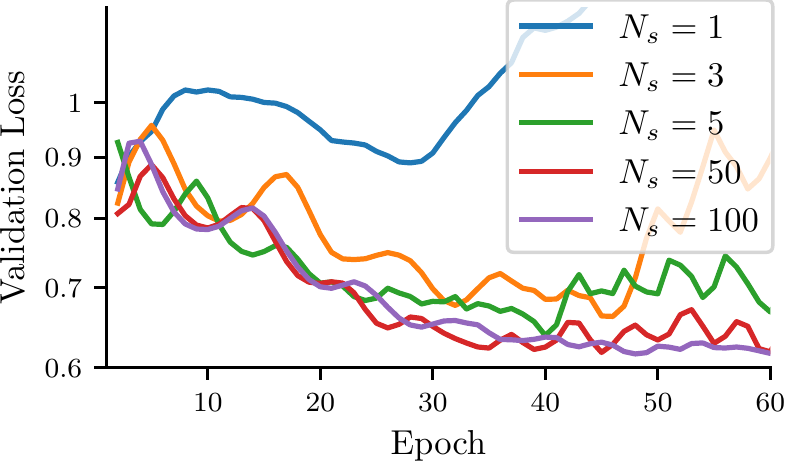}
   \end{center}
   \caption{Evolution of the validation loss for a varying number of samples $N_s$ in the Monte Carlo integration used to calculate the training loss. The validation loss is evaluated using the trapezoidal rule with 20 points. As training set, a simulated dataset with 200 samples is used.}\label{fig:samples}
\end{figure}

In \figref{fig:samples} the results from training the model using a varying number of samples in the Monte Carlo integration are shown. Here it can be seen that using more samples tend to give a better model, but only to a certain level since the result from using 50 and 100 samples are quite similar. But it can also be noted that using only 3 samples results in a model with performance close to that of the best model. However, using a higher number of samples (e.g. 100) gives a smoother validation loss which gives a more robust behavior, in the sense that training multiple models tend to give models with similar performance.

\begin{figure} 
   \begin{center}
      \includegraphics[width=\columnwidth]{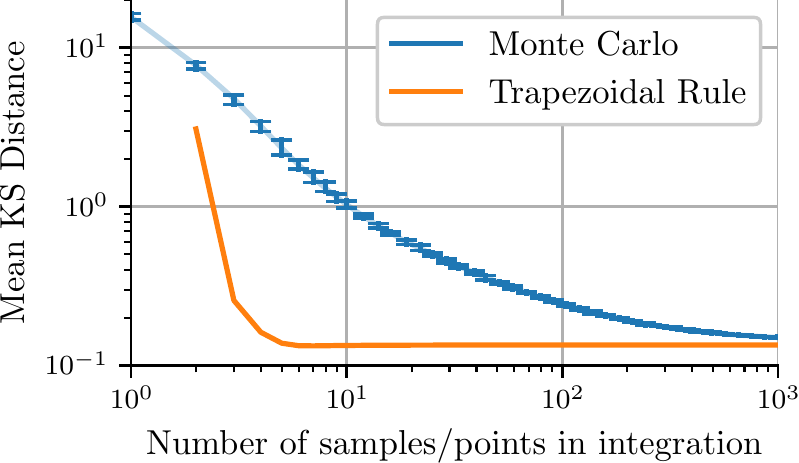}
   \end{center}
   \caption{Mean Kolmogorov–Smirnov Distance for a varying number of points in the integration. The model on which the evaluation is based is trained on 200 samples of the simulated data. Due to the stochastic nature of the Monte Carlo integration, the distance was evaluated 20 times for each point, and the vertical lines indicate the best and worst result for each point. For reference, it can be noted that for the constant model $S_\theta(t,x)=0.5$ it holds that $\barDks^{t_m}\leq 0.5$, with equality as $t_m\rightarrow\infty$.}\label{fig:sampling_prediction}
\end{figure}

In \figref{fig:sampling_prediction} the effect the number of points in the integration has on the accuracy of the predictions given by the model is investigated. As can be seen, the trapezoidal rule is preferable for prediction since it is close to convergence already for around 10 samples, while Monte Carlo integration with 1,000 samples is not. From this it can also be concluded that splitting the integral into two parts, which is what allows the trapezoidal rule to be used, is a key step in creating a practically useful model.

\subsection{Comparison of Models on the Simulated Data}\label{sec:sim_data}
Using the simulated data, the energy-based model is compared with that of the \textit{piecewise constant hazard method} (PCH) and PMF method, both described in \cite{kvammeContinuousDiscretetimeSurvival2021}. Comparisons with the methods \textit{Logistic-Hazards} \citep{gensheimerScalableDiscretetimeSurvival2019,kvammeContinuousDiscretetimeSurvival2021} and MTLR \citep{fotsoDeepNeuralNetworks2018} has also been done, but are not presented here since the results were very similar to that of the PCH and PMF, respectively, which is in line with the results in \cite{kvammeContinuousDiscretetimeSurvival2021}.  

For all models, a feed-forward network consisting of two fully-connected layers with the same number of nodes in each layer, and the rectified linear unit as the activation function, is used. For training, the Adam optimizer implemented in PyTorch is used for all models, and to reduce the risk of overfitting early stopping is used. The hyperparameters for each model can be found in \tabref{tab:sim_hyper}. Notably, the energy-based model seems to benefit from a larger number of nodes in the network, while for the other methods increasing the number of nodes does not improve the results and only increases the risk of overfitting. 
\begin{table}
   
   \caption{Hyperparameters for the simulated data. $N_\text{nodes}$ corresponds to the number of nodes in each layer, and $N_\text{grid}$ is the number of grid points in the discretization used in the PCH and PMF method.  }
   \label{tab:sim_hyper}
   \center
   
   \begin{tabular}{c c c c c c c}
     \hline
     \
     &  & &\multicolumn{2}{c}{\textbf{200 Samples}}
      & \multicolumn{2}{c}{\textbf{1,000 Samples}}  \\ \cmidrule(r){4-5} \cmidrule(l){6-7 }
      Model & $N_\text{nodes}$ & Dropout& L. Rate & $N_\text{bin}$ & L. Rate &  $N_\text{grid}$ \\
                \hline
     EBM &    64 &    0    &    0.02 &    -- &    0.02 &   -- \\
     PCH &    32 &    20 \% &    0.005&    5 &    0.005 &    15 \\
     PMF &    32 &    20 \% &    0.005 &    5 &    0.005 &    15\\
     \hline
   \end{tabular}
\end{table}
\begin{figure}[ht]
   \begin{center}
      \includegraphics[width=\columnwidth]{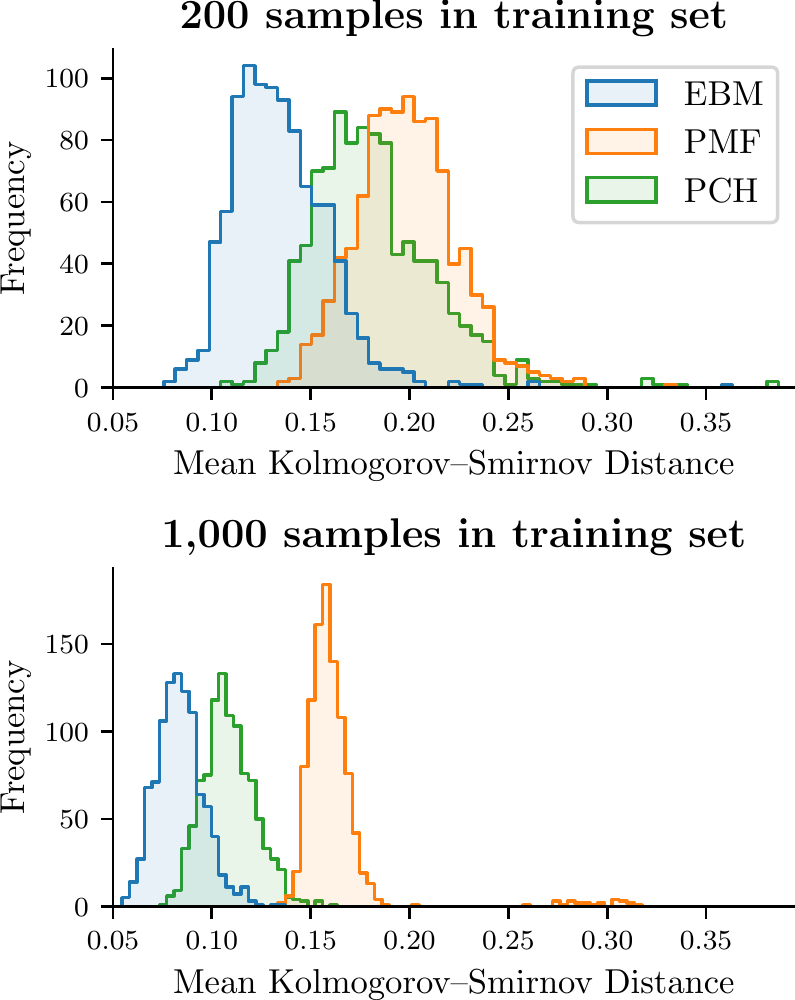}
   \end{center}
   \caption{Results from a Monte Carlo study based on 1,000 different representations of the datasets. }
   \label{fig:MC_hist}
\end{figure}

\begin{figure}[ht]
   \begin{center}
      \includegraphics[width=\columnwidth]{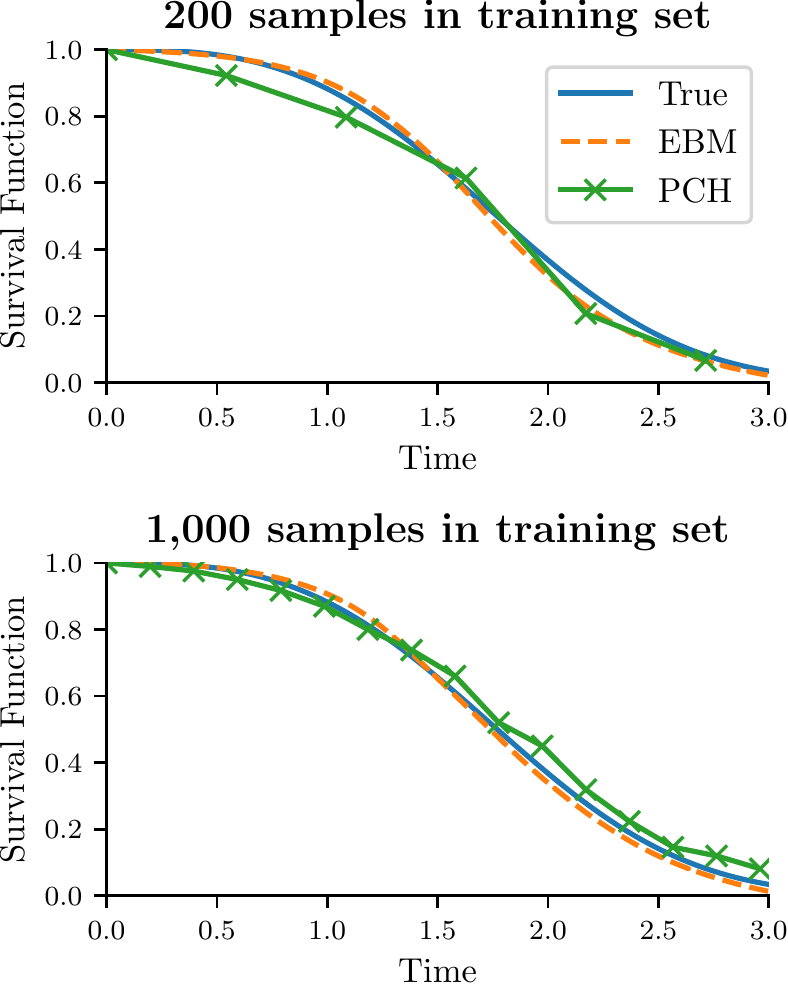}
   \end{center}
   \caption{Predicted survival functions from the energy-based model and the PCH model compared with the true survival function for $k=3$ and $\lambda=2$. }
   \label{fig:sim_example}
\end{figure}

A comparison of the models, based on a Monte Carlo study of 1,000 different representations of the datasets, can be found in \figref{fig:MC_hist}. As can be seen, the energy-based model shows significantly better performance compared to the other models on average, and worst-case performance similar to the average performance of the other models. In \figref{fig:sim_example} the predicted survival functions from the energy-based model, and the PCH model from one of the experiments are compared with the true survival function. Notable is the smooth appearance of the energy-based model, which is a characteristic of this model.

\subsection{Predictive Maintenance of Starter Batteries}
Here the performance of the model on experimental data is investigated using the vehicle fleet data. For comparison, the models from \secref{sec:sim_data} are also included. The same architecture and training procedure are also used, but now with the hyperparameters in \tabref{tab:exp_hyper}. 

Since the true distribution is now unknown, the comparison is based on receiver operating characteristic (ROC) curves from using the models in the predictive maintenance scheme. For each model, and a given threshold, a point in the ROC curve is calculated in the following way: if, based on the predictions given by the model, the decision is to replace the battery and the recorded failure time of the battery is within the considered horizon, the decision is marked as a \textit{true positive}, and if it does not fail it is marked as a \textit{false positive}. Using this the ROC curve is created by performing a sweep in the threshold. 
\begin{figure*}[h]
   \begin{center}
      \includegraphics[width=\textwidth]{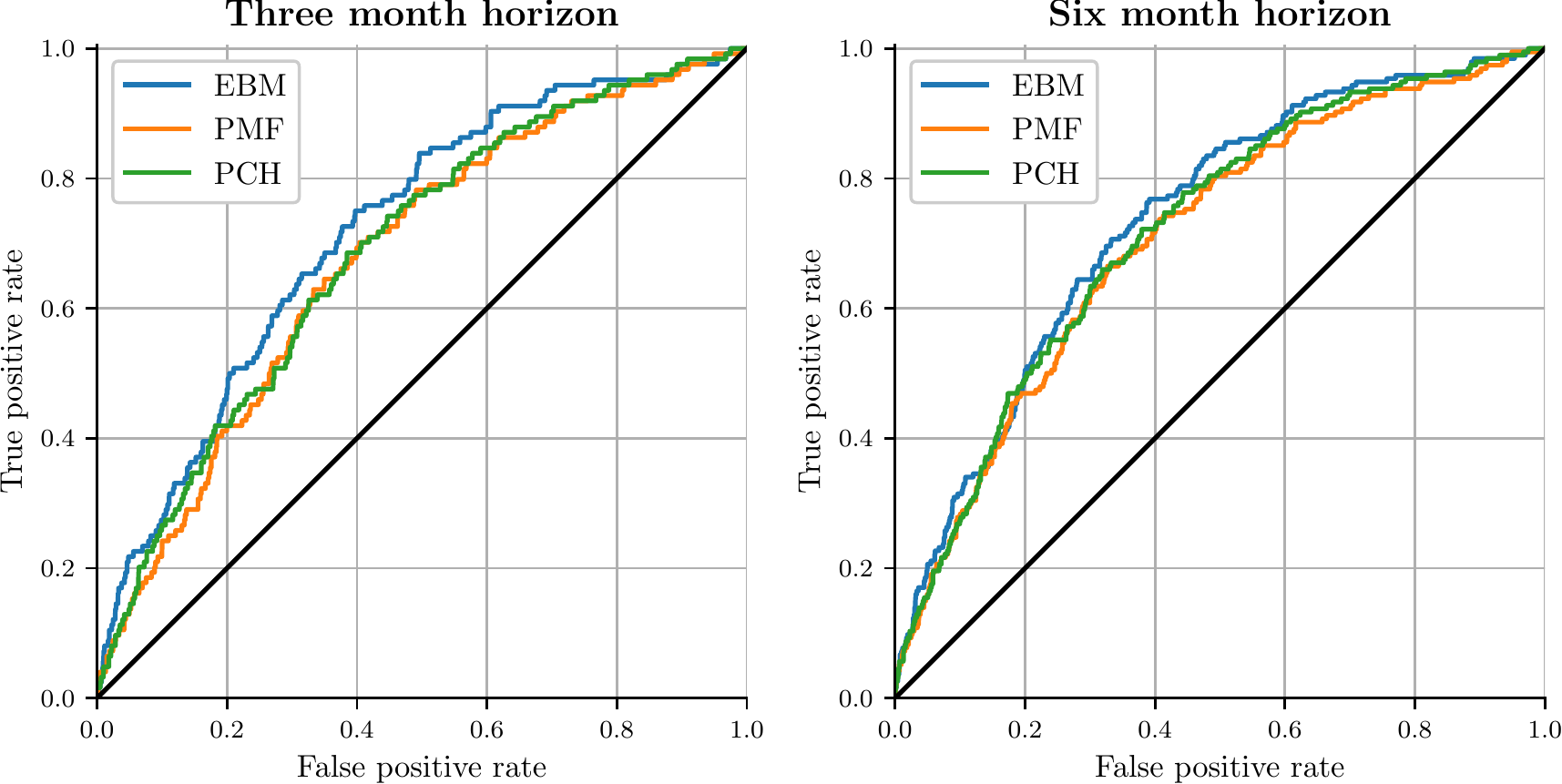}
   \end{center}
   \caption{ROC curves for prediction horizons of three and six months for all models, with the identity line in black as a reference.}\label{fig:ROC}
\end{figure*}
In \figref{fig:ROC} the ROC curves for time horizons of three and six months are shown. Here the energy-based model again shows the best performance by being closest to the upper left corner in most parts, which is the ideal value. However, the results are not as clear as for the simulated data, especially for the longer time horizon. Here it should also be considered that the ROC curve of the true distribution is unknown, and therefore we do not know how the curves of the models relate to it. For example, if the ROC curve of the energy-based model is very close to that of the true distribution, one could argue that it is significantly better than the other models. But on the other hand, if it is much worse than that of the true distribution, a more correct conclusion would probably be that the performances of the models are quite similar.

\begin{table}

   \caption{Hyperparameters for the vehicle fleet data. $N_\text{nodes}$ corresponds to the number of nodes in each layer, and $N_\text{grid}$ is the number of grid points in the discretization used in the PCH and PMF method.}
   \center
   \begin{tabular}{c c c c c }
     \hline
     Model & $N_\text{nodes}$ & Dropout& L. Rate & $N_\text{bin}$  \\
                \hline
     EBM &    400 &    20 \% &    0.01 &    --    \\
     PCH &    200 &    20 \% &    0.001&    20    \\
     PMF &    200 &    20 \% &    0.001 &    20    \\
     \hline
   \end{tabular}
   \label{tab:exp_hyper}
\end{table}
\
\section{Conclusion}
In this paper, energy-based modeling is applied to the problem of survival modeling. A key step in this is the introduction of right-censored data, which, based on a maximum likelihood approach, is shown to be a straightforward process. 

An important part of the model is the evaluation of the integral used in the normalization of the probability density. By splitting the integral into two parts, one corresponding to the tail of the distribution whose integral is calculated directly based on a single point, and the other being the remaining proper integral which can be evaluated using the trapezoidal rule,  it is shown that predicting the survival function can be done efficiently using only a few evaluations of the network. During training, however, a sampling-based method is still required. 

The model is compared with existing models using both simulated data and experimental data from a fleet of vehicles, and the results indicate a highly competitive performance of the model. This is especially clear in the case of the simulated data.



\bibliography{bibliography_BBT}             

\begin{thebibliography}{18}
\providecommand{\natexlab}[1]{#1}
\providecommand{\url}[1]{\texttt{#1}}
\providecommand{\urlprefix}{URL }
\expandafter\ifx\csname urlstyle\endcsname\relax
  \providecommand{\doi}[1]{doi:\discretionary{}{}{}#1}\else
  \providecommand{\doi}{doi:\discretionary{}{}{}\begingroup
  \urlstyle{rm}\Url}\fi

\bibitem[{Bengio et~al.(2000)Bengio, Ducharme, and
  Vincent}]{bengioNeuralProbabilisticLanguage2000}
Bengio, Y., Ducharme, R., and Vincent, P. (2000).
\newblock A neural probabilistic language model.
\newblock \emph{Advances in neural information processing systems}, 13.

\bibitem[{Biganzoli et~al.(1998)Biganzoli, Boracchi, Mariani, and
  Marubini}]{biganzoliFeedForwardNeural1998}
Biganzoli, E., Boracchi, P., Mariani, L., and Marubini, E. (1998).
\newblock Feed forward neural networks for the analysis of censored survival
  data: A partial logistic regression approach.
\newblock \emph{Statistics in Medicine}, 17(10), 1169--1186.
\newblock
  \doi{10.1002/(SICI)1097-0258(19980530)17:10<1169::AID-SIM796>3.0.CO;2-D}.

\bibitem[{Brown et~al.(1997)Brown, Branford, and
  Moran}]{brownUseArtificialNeural1997}
Brown, S., Branford, A., and Moran, W. (1997).
\newblock On the use of artificial neural networks for the analysis of survival
  data.
\newblock \emph{IEEE Transactions on Neural Networks}, 8(5), 1071--1077.
\newblock \doi{10.1109/72.623209}.

\bibitem[{Ching et~al.(2018)Ching, Zhu, and
  Garmire}]{chingCoxnnetArtificialNeural2018}
Ching, T., Zhu, X., and Garmire, L.X. (2018).
\newblock Cox-nnet: {{An}} artificial neural network method for prognosis
  prediction of high-throughput omics data.
\newblock \emph{PLOS Computational Biology}, 14(4), e1006076.
\newblock \doi{10.1371/journal.pcbi.1006076}.

\bibitem[{Cox(1972)}]{coxAnalysisMultivariateBinary1972}
Cox, D.R. (1972).
\newblock The {{Analysis}} of {{Multivariate Binary Data}}.
\newblock \emph{Journal of the Royal Statistical Society. Series C (Applied
  Statistics)}, 21(2), 113--120.
\newblock \doi{10.2307/2346482}.

\bibitem[{Dhada et~al.(2022)Dhada, Parlikad, Steinert, and
  Lindgren}]{dhadaWeibullRecurrentNeural2022}
Dhada, M., Parlikad, A.K., Steinert, O., and Lindgren, T. (2022).
\newblock Weibull recurrent neural networks for failure prognosis using
  histogram data.
\newblock \emph{Neural Computing and Applications}.
\newblock \doi{10.1007/s00521-022-07667-7}.

\bibitem[{Du and Mordatch(2019)}]{duImplicitGenerationModeling2019}
Du, Y. and Mordatch, I. (2019).
\newblock Implicit generation and modeling with energy based models.
\newblock \emph{Advances in Neural Information Processing Systems}, 32.

\bibitem[{Fotso(2018)}]{fotsoDeepNeuralNetworks2018}
Fotso, S. (2018).
\newblock Deep {{Neural Networks}} for {{Survival Analysis Based}} on a
  {{Multi-Task Framework}}.

\bibitem[{Gao et~al.(2018)Gao, Lu, Zhou, Zhu, and
  Wu}]{gaoLearningGenerativeConvnets2018}
Gao, R., Lu, Y., Zhou, J., Zhu, S.C., and Wu, Y.N. (2018).
\newblock Learning generative convnets via multi-grid modeling and sampling.
\newblock In \emph{Proceedings of the {{IEEE Conference}} on {{Computer
  Vision}} and {{Pattern Recognition}}}, 9155--9164.

\bibitem[{Gensheimer and
  Narasimhan(2019)}]{gensheimerScalableDiscretetimeSurvival2019}
Gensheimer, M.F. and Narasimhan, B. (2019).
\newblock A scalable discrete-time survival model for neural networks.
\newblock \emph{PeerJ}, 7, e6257.
\newblock \doi{10.7717/peerj.6257}.

\bibitem[{Gustafsson et~al.(2020)Gustafsson, Danelljan, Timofte, and
  Sch{\"o}n}]{gustafssonHowTrainYour2020}
Gustafsson, F.K., Danelljan, M., Timofte, R., and Sch{\"o}n, T.B. (2020).
\newblock How to train your energy-based model for regression.
\newblock \emph{arXiv preprint arXiv:2005.01698}.

\bibitem[{Ishwaran et~al.(2008)Ishwaran, Kogalur, Blackstone, and
  Lauer}]{ishwaranRandomSurvivalForests2008}
Ishwaran, H., Kogalur, U.B., Blackstone, E.H., and Lauer, M.S. (2008).
\newblock Random survival forests.
\newblock \emph{The Annals of Applied Statistics}, 2(3), 841--860.
\newblock \doi{10.1214/08-AOAS169}.

\bibitem[{Katzman et~al.(2018)Katzman, Shaham, Cloninger, Bates, Jiang, and
  Kluger}]{katzmanDeepSurvPersonalizedTreatment2018}
Katzman, J.L., Shaham, U., Cloninger, A., Bates, J., Jiang, T., and Kluger, Y.
  (2018).
\newblock {{DeepSurv}}: Personalized treatment recommender system using a
  {{Cox}} proportional hazards deep neural network.
\newblock \emph{BMC Medical Research Methodology}, 18(1), 24.
\newblock \doi{10.1186/s12874-018-0482-1}.

\bibitem[{Kvamme and Borgan(2021)}]{kvammeContinuousDiscretetimeSurvival2021}
Kvamme, H. and Borgan, {\O}. (2021).
\newblock Continuous and discrete-time survival prediction with neural
  networks.
\newblock \emph{Lifetime Data Analysis}, 27(4), 710--736.
\newblock \doi{10.1007/s10985-021-09532-6}.

\bibitem[{LeCun et~al.()LeCun, Chopra, Hadsell, Ranzato, and
  Huang}]{lecunTutorialEnergyBasedLearning}
LeCun, Y., Chopra, S., Hadsell, R., Ranzato, M., and Huang, F.J. (????).
\newblock A {{Tutorial}} on {{Energy-Based Learning}}.
\newblock \emph{MIT Press}, 59.

\bibitem[{Li et~al.(2022)Li, Krivtsov, and
  Arora}]{liAttentionbasedDeepSurvival2022}
Li, X., Krivtsov, V., and Arora, K. (2022).
\newblock Attention-based deep survival model for time series data.
\newblock \emph{Reliability Engineering \& System Safety}, 217, 108033.
\newblock \doi{10.1016/j.ress.2021.108033}.

\bibitem[{Song and Kingma(2021)}]{songHowTrainYour2021}
Song, Y. and Kingma, D.P. (2021).
\newblock How to train your energy-based models.
\newblock \emph{arXiv preprint arXiv:2101.03288}.

\bibitem[{Voronov et~al.(2020)Voronov, Krysander, and
  Frisk}]{voronovPredictiveMaintenanceLeadacid2020}
Voronov, S., Krysander, M., and Frisk, E. (2020).
\newblock Predictive maintenance of lead-acid batteries with sparse vehicle
  operational data.
\newblock \emph{International Journal of Prognostics and Health Management},
  11(1).

\end{thebibliography}


\end{document}